\documentclass[12,final]{elsarticle}

\usepackage{geometry}
\usepackage{graphicx,amsmath,amssymb,amsfonts}   
\usepackage{dcolumn}    
\usepackage{bm}         
\usepackage{color}
\usepackage{times,setspace,multirow}
\usepackage{lineno}

\newcommand{\ie}{\emph{i.e.\@} }

\newcolumntype{d}[1]{D{.}{.}{#1}}

\begin{document}
\title{Scoping of material response under DEMO neutron irradiation:
comparison with fission and influence of nuclear library selection
}


\author[ccfe]{M. R. Gilbert\corref{cor1}}
\author[ccfe]{J.-Ch. Sublet}
\address[ccfe]{Culham Centre for Fusion Energy, Culham Science Centre, Abingdon, OX14 3DB, UK}

\cortext[cor1]{Corresponding author, mark.gilbert@ccfe.ac.uk}

\begin{abstract}
Predictions of material activation inventories will be a key input to virtually all aspects of the operation, safety and environmental assessment of future fusion nuclear plants. Additionally, the neutron-induced transmutation (change) of material composition (inventory) with time, and the creation and evolution of configurational damage from atomic displacements, require precise quantification because they can lead to significant changes in material properties, and thus influence reactor-component lifetime.
A comprehensive scoping study has been performed to quantify the activation, transmutation (depletion and build-up) and immediate damage response under neutron irradiation for all naturally occurring elements from hydrogen to bismuth. The resulting database provides a global picture of the response of a material, covering the majority of nuclear technological space, but focussing specifically on typical conditions expected for a demonstration fusion power plant (DEMO).

Results from fusion are compared against typical fission conditions for selected fusion relevant materials, demonstrating that the latter cannot be relied upon to give accurate scalable experimental predictions of material response in a future fusion reactor. Results from different nuclear data libraries are also compared, highlighting the variations and deficiencies.

\end{abstract}


\maketitle
\section{Introduction}

Accurate predictions of material activation -- as caused by nuclear reactions -- are necessary and vital because they determine the operational limits, maintenance schedule, safety and environmental impact of future fusion nuclear plants. Meanwhile, the resultant neutron-induced transmutation of material composition (inventory), including the formation of gas bubbles, and the creation and evolution of damage from atomic displacements, require precise quantification because they can lead to a change in material properties, which influence reactor-component lifetimes and hence reactor maintenance costs.

For a fully-detailed reactor design with complete materials descriptions, such as that now available for neutron-transport simulations of the ITER experimental reactor currently under construction, it is entirely possible to accurately predict the neutron irradiation at any point in the vessel. Large-scale parallel computing with modern neutron-transport codes, such as MCNP-6~\cite{mcnp1}, can handle the hundreds of thousands of finite-element cells required to describe the reactor geometry. Subsequently, these predictions of neutron fields can be fed into a nuclear inventory code, such as the state-of-the-art  FISPACT-II~\cite{subletetal2015} system, to calculate the material activation and transmutation response for any reactor component at any given time during reactor operation or after shutdown.

However, the design of a demonstration fusion power plant (DEMO) is still at the conceptual stage, and it is far too early to build detailed reactor geometries, with many of the material choices still uncertain. Indeed, there is and should be plenty of scope for material selection and design studies, with almost no limit, in principle, on what materials (elements) can be used. There are, of course, many factors to consider when deciding what material composition might be acceptable for a particular DEMO application, and foremost amongst these is the activation response. A material's compositional stability under neutron irradiation -- \ie how much will it transmute? --, particularly how much gases and other harmful impurities it might generate, is also a chief concern, as is the associated formation of radiation damage, whose source terms (the primary knock-on atoms or PKAs) can be quantified using the nuclear interaction database~\cite{gilbertmariansublet2015}.

To aid the material selection decisions of nuclear engineers and scientists, and also material modellers and experimentalists, there is a need for a reference document of response measures for elements under typical fusion (and fission) conditions. Such documents have been produced before, but these are now obsolete because they do not use the latest, modern, and more complete nuclear reaction data and simulation tools. In fact nuclear data libraries, such as the Talys Evaluated Nuclear Data Libraries (TENDL), of which TENDL-2014~\cite{tendl2014} is the most recent, are constantly changing, and so it is important that such scoping documents be easily and quickly reproducible -- then they can be released shortly after the libraries themselves. Thus, significant effort has been expended in creating a largely automated computational platform, capable of performing the tens of thousands inventory simulations required, and collating them into a single report. Note that the conceptual designs of (DEMO) fusion reactors are also changing on a regular basis, which is an additional incentive for making the creation of such exploratory reports rapid -- to make them applicable to the most recent engineering concepts for fusion power.

This paper describes a comprehensive scoping study of the activation, transmutation and primary damage (PKA) characteristics of all naturally occurring elements from hydrogen to bismuth performed with FISPACT-II~\cite{subletetal2015} and TENDL-2014~\cite{tendl2014}, together with a modern nuclear decay data file, UKDD-12 (built from EAF-2007 decay data~\cite{forrest2007} with inclusion of some
updates and an increased set of short-lived nuclides to cover the advancements in TENDL). These results, already published in an openly available report~\cite{fusionhandbook2015}, provide a global picture of the response of a material, covering the majority of nuclear technological space, but focussing specifically on typical conditions expected for a DEMO fusion power plant.

An important aspect to the scoping calculations is the ability to explore trends in particular nuclear observables, such as gas production rates, or shutdown activities. Thus, since it is not practical to present anything other than examples of the results presented in the report~\cite{fusionhandbook2015}, here, instead, we focus on the comparison of certain DEMO-TENDL-2014 results for a fusion-relevant subset of the elements. Furthermore, to highlight variations and deficiencies, we also contrast the TENDL-2014 results with those calculated using other international nuclear cross section libraries: ENDF/BVII.1~\cite{endfb72011}, JEFF-3.2~\cite{JEFF32}, and JENDL-4.0~\cite{jendl4.0}. Additional reports~\cite{pwrfissionhandbook,hfrfissionhandbook,fbrfissionhandbook} have also been produced for typical fission scenarios, and results from these are also discussed and compared against fusion equivalents.

\section{Calculations and results per element}\label{calculations}

\begin{figure}[h]
{\includegraphics[width=1.0\textwidth]
{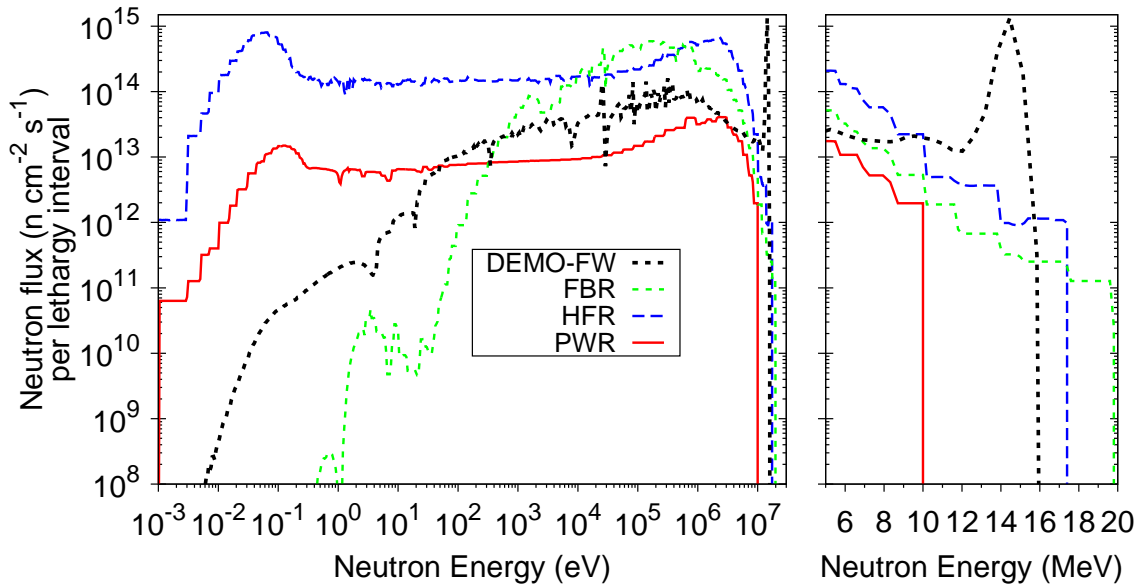}}
\caption{\label{neutron_spectra}(colour online) The four neutron spectra used for the scoping calculations, where DEMO-FW is a typical first wall spectrum for a demonstration fusion power plant (total integrated flux of \(5.9\times10^{14}\)~n~cm\(^{-2}\)~s\(^{-1}\)); FBR is the core assembly spectrum for the large-scale prototype fast breeder superphenix reactor that was located in the south of France (\(2.4\times10^{15}\)~n~cm\(^{-2}\)~s\(^{-1}\)); HFR is the spectrum for volume-averaged low-flux material test location of the high-flux reactor at Petten, Netherlands (\(5.3\times10^{14}\)~n~cm\(^{-2}\)~s\(^{-1}\)); and PWR is the fuel assembly-averaged spectrum for the type P4 pressurized-water reactor at the Paluel site in France (\(3.25\times10^{14}\)~n~cm\(^{-2}\)~s\(^{-1}\)). Left: the spectra on a logarithmic eV scale showing
the full energy range. Right: a linear MeV scale, showing the high-energy parts of the
spectra, in particular the fission tails.}
\end{figure}

Figure~\ref{neutron_spectra} shows the primary DEMO first wall (FW) neutron spectrum used for the calculations in~\cite{fusionhandbook2015}, together with the three fission spectra considered for the additional TENDL-2014-based reports already mentioned. An automated script runs, for each element, a set of irradiations  with this spectrum (followed by cooling, as appropriate) using the FISPACT-II inventory code, and produces the following outputs, which are fully described in~\cite{fusionhandbook2015}:

\begin{enumerate}
\item \underline{Activation tables} -- six tables, one for each of total (specific) activity (Bq~kg\(^{-1}\)), decay heat (kW~kg\(^{-1}\)), \(\gamma\) (contact) dose rate (Sv~h\(^{-1}\)), inhalation and ingestion dose (Sv~h\(^{-1}\)), and (IAEA) clearance index, showing the percentage contributions to the particular  radiological response quantity at a range of cooling times following a 2 full-power-year (fpy) irradiation in DEMO FW spectrum. The total radiological response is also given at each cooling time.

\item \underline{Activation graphs} -- three pairs of tables, one for each of total activity, decay heat and \(\gamma\) dose rate. One graph shows the evolution in response of the material during cooling after 2~fpy irradiations in the DEMO-FW environment and two other typical in-vessel DEMO conditions. The second graph shows the response cooling following the FW irradiation, but also includes an indication of which radionuclides are dominant for the radiological quantity at a particular cooling time. Nuclides appear in such plots at the position on the time (x) axis corresponding to their half-life and on the activity (y) axis corresponding to their contribution to the radiological quantity at shutdown after the 2 fpy irradiation.

\item \underline{Importance diagrams}~\cite{forrest1998,gilbertetalNSE2014} -- plots showing the regions of the incident-neutron-energy versus decay-time landscape where a single radionuclide dominates (contributes more than 50\%) a particular activation quantity; providing a general visual representation of the post-irradiation response of a material that is independent of the neutron spectrum. Diagrams for total activity, decay heat, and \(\gamma\) dose rate are produced from a sequence of 2~fpy irradiations under mono-energetic neutron spectra (a high resolution 709-group neutron energy group structure was used to divide up the energy range -- see~\cite{subletetal2015}) at a characteristic fusion neutron flux of 10\(^{15}\)~n~cm\(^{−2}\)~s\(^{-1}\).

\item \underline{Transmutation results} -- the burn-up response, where a material's composition changes with time due to irradiation, is calculated during a 2~fpy DEMO FW irradiation and then plotted to show the growth of impurity elements, including gases He and H, and corresponding depletion of the parent element. The He and H production in atomic parts per million (appm) in 1 fpy, and an estimate of the displacements per atom (dpa) per fpy, are output explicitly (as text). The final nuclide composition after 2~fpy is also presented in a ``chart of the nuclides''~\cite{gilbertetalNSE2014} tableau.

\item \underline{PKA distributions} -- primary knock-on atoms (PKAs) are the radiation damage source terms that determine the size of displacement damage cascades and hence the population of structural defects created in irradiated materials. They are important inputs to both atomistic simulations and the design of irradiation experiments. A newly written code, SPECTRA-PKA~\cite{gilbertmariansublet2015}, combines the DEMO FW neutron spectrum with neutron energy versus PKA recoil energy cross section (probability) matrices to define PKA-energy distributions for both nuclide (isotope) daughters and elemental recoil sums (more useful for atomistic modelling where different isotopes of the same element cannot normally be distinguished), which are then plotted in separate graphs.

\item \underline{Pathway analysis} -- the important production pathways are calculated and output for every radionuclide considered significant because it appears in either the importance
diagrams or activation tables of the element. A standard FISPACT-II pathway analysis is performed for 2-fpy 10\(^{15}\)~n~cm\(^{−2}\)~s\(^{-1}\) irradiations at four neutron energy ranges: thermal neutrons from 0.02 to 0.05 eV; intermediate neutrons from 20 to 40 keV; fast neutrons at typical fission energies of 1 to 3 MeV; and a fusion relevant range from 13 to 15 MeV. The output nuclides are listed in order of increasing decay half-life.

\end{enumerate}

As an example of the typical output for an element, the complete set of results for Fe are given in the supplementary data. For equivalent reports under typical fission reactor conditions~\cite{pwrfissionhandbook,hfrfissionhandbook,fbrfissionhandbook}, the same calculations, as described above, are repeated but with the DEMO neutron spectrum replaced with the appropriate fission spectrum (see figure~\ref{neutron_spectra}). Note that the importance diagrams do not need repeating for reports that use the same nuclear data libraries as they are spectrum independent and thus identical in all such reports. This is fortuitous because the importance diagrams alone require almost 60,000 separate FISPACT-II calculations, while the remaining calculations across all elements need less than a further 1000 calculations per report.

\section{Elemental trends: variation with neutron field}

\begin{figure}[h]
{\includegraphics[width=1.0\textwidth]
{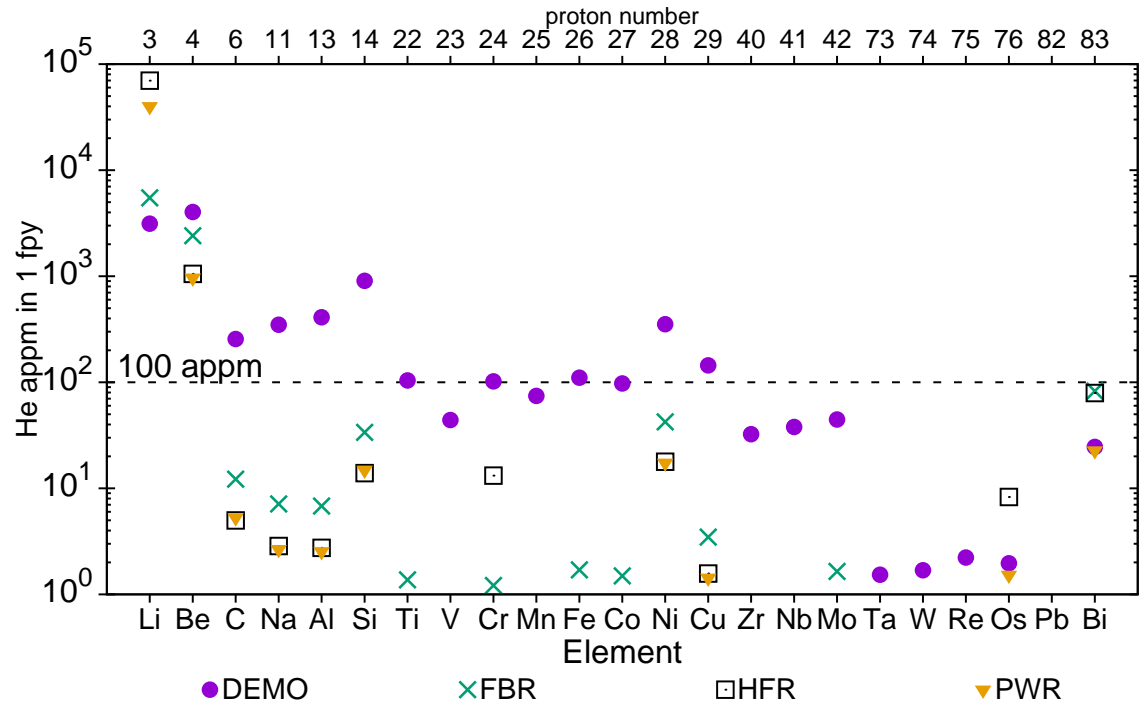}}
\caption{\label{heproductionenviro}(colour online) Variation in He production (in atomic parts per million or appm) per fpy as a function of element and irradiation environment. All calculations performed using FISPACT-II~\cite{subletetal2015} and the TENDL-2014~\cite{tendl2014} nuclear cross section library. The data shown in the plot is also available in tabulated form in the supplementary data.}
\end{figure}

The amount of information produced by the scoping calculations is vast, but the results per element are relatively simple to examine and understand. However, there are several quantities that can be examined at a global level, and some of these can be used to identify elements that satisfy certain limits.
For example, the gas production rates are an important quantity that will influence how much a material will embrittle and swell under irradiation. Figure~\ref{heproductionenviro}, shows the predicted helium production (in appm) during the first 1~fpy of irradiation in the DEMO-FW and three fission spectra for a selection of fusion (and fission) relevant elements (summary plots covering all 81 elements are too complex to be meaningful).

Of the elements considered in figure~\ref{heproductionenviro}, Li (vital for tritium breeding in fusion) produces the largest helium production for the three fission systems, while Be (an important neutron multiplier) is the highest in the fusion environment due to the increased occurrence of the threshold (n,\(\alpha\)) reaction on \(^{9}\)Be. The (n,t) on \(^{6}\)Li, on the other hand, which produces He as the daughter, is non-threshold, and thus the fission spectra, with higher thermal neutron fluxes, produce the greatest He production from Li. Note in figure~\ref{heproductionenviro}, and in subsequent plots in this paper, that some elements have missing points because they produce less than the minimum of the y-axis scale -- as is the case for the row-three transition metals, Ta, W, Re, which show a very low \(\sim\)2~appm He in 1~DEMO-fpy, but no values for the three fission scenarios because they are less than 1~appm. Notice also Pb, where none of the four scenarios produce He at a rate above 1~appm.

In general, the He production in the DEMO environment is higher for most elements, again because of the increase in threshold (n,\(\alpha\)) reactions, and, furthermore, production decreases with increasing mass. For example, in Fe, which is the main component in steels, He production is around 100~appm under DEMO conditions, but 1~appm or less per fpy in the three fission environments. An even greater disparity is observed for Al, where the fusion environment is predicted to produce more than 400~appm in a single fpy, but less than 10~appm are predicted in fission.

\begin{figure}[h]
{\includegraphics[width=1.0\textwidth]
{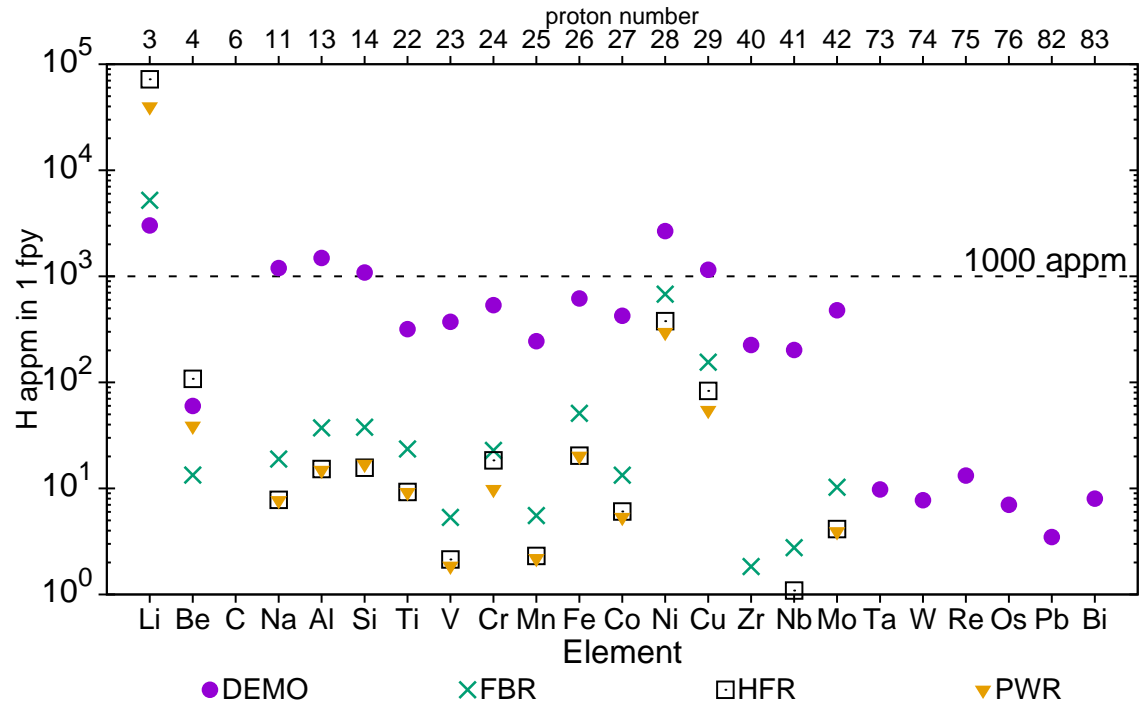}}
\caption{\label{hproductionenviro}(colour online) Variation in H appm per fpy as a function of element and irradiation environment. TENDL-2014~\cite{tendl2014} results. The data shown in the plot is also available in tabulated form in the supplementary data.}
\end{figure}

Equivalently, figure~\ref{hproductionenviro}, shows hydrogen production as a function of element and irradiation scenario. As with He production (figure~\ref{heproductionenviro}), H production is significantly enhanced for most elements in a fusion environment -- by at least an order of magnitude in metals --, which demonstrates that behavioural predictions for materials based on experience in fission reactors, may not be reliable, particularly with regard to swelling and embrittlement. Al once again shows one of the largest differences, ranging from almost 1500~appm H in DEMO, to less than 50 in all three fission systems (see the supplementary data for exact values).

Note also here the different scale in figure~\ref{heproductionenviro} for H compared to figure~\ref{heproductionenviro} for He (as illustrated by the horizontal lines in each plot representing the order of magnitude of production for the first row of transition metals under DEMO-FW conditions -- they are an order of magnitude different in the two plots). Hydrogen production is significantly higher than helium production for most elements.

\begin{figure}[h]
{\includegraphics[width=1.0\textwidth]
{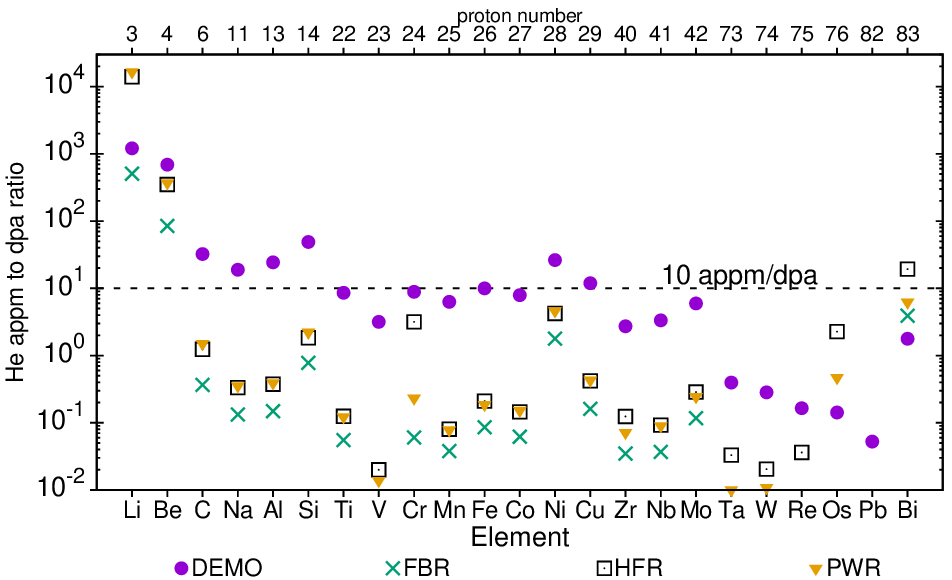}}
\caption{\label{hetodpaenviro}(colour online) Variation in He appm production to dpa as a function of element and irradiation environment. TENDL-2014~\cite{tendl2014} results. The data shown in the plot is also available in tabulated form in the supplementary data.}
\end{figure}

\begin{figure}[h]
{\includegraphics[width=1.0\textwidth]
{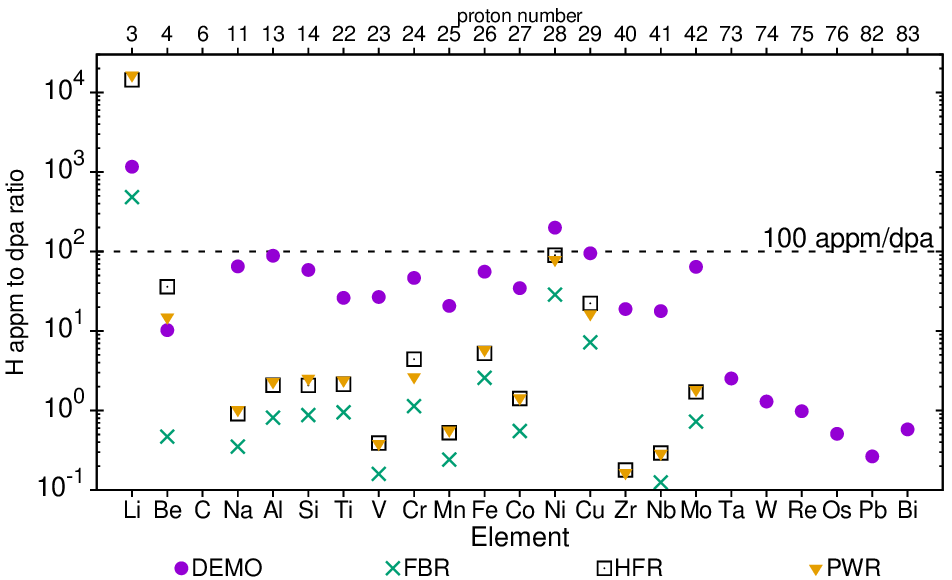}}
\caption{\label{htodpaenviro}(colour online) Variation in H appm production to dpa as a function of element and irradiation environment. TENDL-2014~\cite{tendl2014} results. The data shown in the plot is also available in tabulated form in the supplementary data.}
\end{figure}

This difference between fusion and fission gas production is further emphasized by figures~\ref{hetodpaenviro} and~\ref{htodpaenviro}, which show, respectively, the He- and H-to-dpa ratios for the same elements and irradiation conditions.
These figures suggest that much higher damage doses would need to be achieved in a fission experiment (such as HFR) to attain the correct (for fusion) levels of gas production. Note that the dpa values used here were obtained using the NRT~\cite{norgettetal1975} formula, with standard threshold displacement energy \(E_d\) values for some elements from Greenwood and Smither (table I in~\cite{greenwoodsmither1981}, or reproduced as table II in~\cite{macfarlanekahler2010}), with the exception of W where a more realistic \(E_d=53\)~eV was used, and 25~eV otherwise~\cite{subletetal2015}.

\begin{figure}[h]
{\includegraphics[width=1.0\textwidth]
{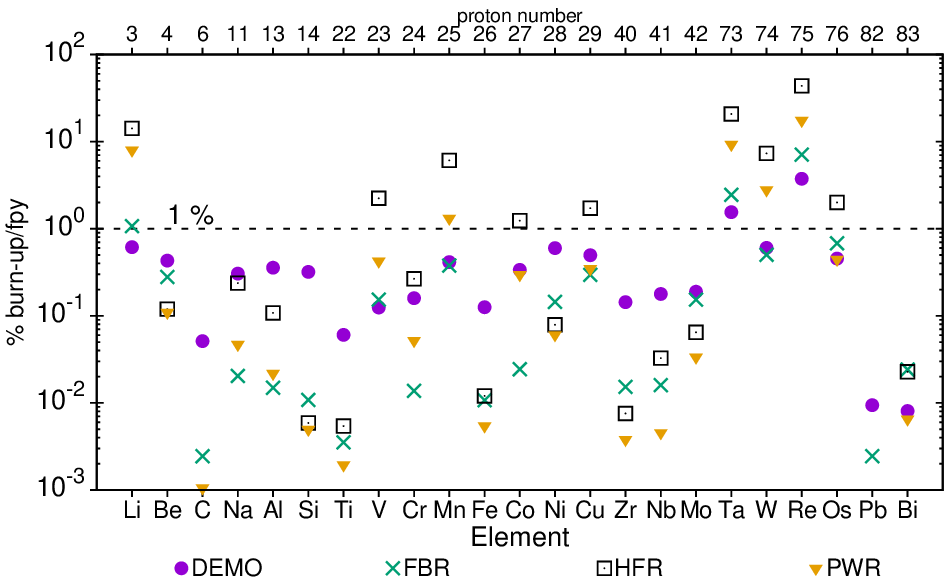}}
\caption{\label{burnupenviro}(colour online) Variation in burn-up transmutation rates of the parent element under neutron irradiation as a function of element and reactor type. TENDL-2014~\cite{tendl2014} results. The data shown in the plot is also available in tabulated form in the supplementary data.}
\end{figure}

Another important consideration in materials design is the stability of a material's composition under irradiation -- if a material has been carefully designed and manufactured to meet some operational requirement, then it is important that the impurities generated under neutron irradiation do not detrimentally influence performance. One measure of this build-up of impurities is through the change in the amount (burn-up) of the original parent element. Figure~\ref{burnupenviro} shows the \% burn-up per fpy in the selected elements under the four reactor conditions. Under fusion conditions, the row-three transition metals show significant burn-up, even with partial self-shielding correction factors~\cite{gilbertsublet2011}. Notice that the fission reactors typically show greater burn-up rates because of their ``softer'' (more-moderated) neutron spectra, which enhance the neutron capture reactions that eventually lead to the production of higher proton-number elements.

\section{Elemental trends: variation with nuclear library}

The previous section considered how sensitive inventory calculations are to the different neutron fields that elements could be exposed to in real (or planned) nuclear reactors. However, another source of uncertainty and variation comes about due to the availability of different nuclear data libraries, each of which will have been developed for a specific application and may have somewhat differing computational approaches. A key benefit of the automated approach developed for these scoping studies is the ability to easily switch to an alternative nuclear data library (FISPACT-II can read any library in the ENDF-6~\cite{endf605} format).
Calculations with the DEMO-FW spectrum have been repeated for three other international nuclear cross section libraries: ENDF/BVII.1~\cite{endfb72011}, JEFF-3.2~\cite{JEFF32}, and JENDL-4.0~\cite{jendl4.0}.

\begin{figure}[h]
{\includegraphics[width=1.0\textwidth]
{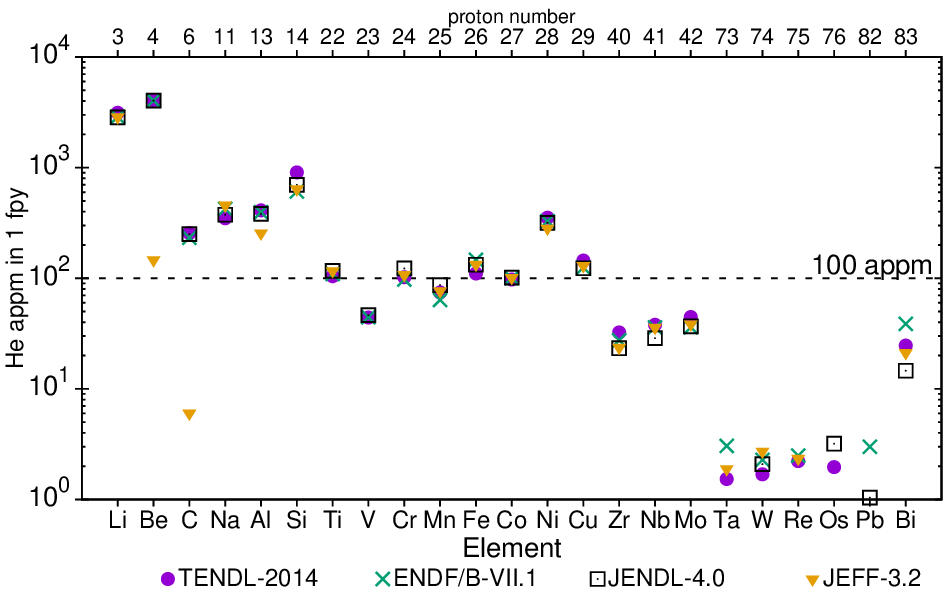}}
\caption{\label{heproductionlib}(colour online) Variation in He production per fpy under the DEMO-FW spectrum as a function of element and choice of nuclear data library. The data shown in the plot is also available in tabulated form in the supplementary data.}
\end{figure}
\begin{figure}[h]
{\includegraphics[width=1.0\textwidth]
{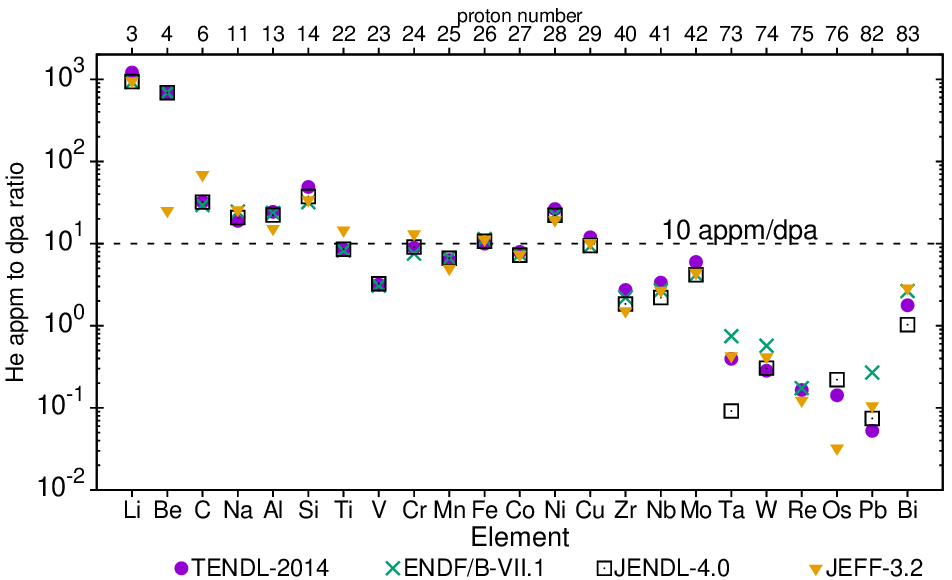}}
\caption{\label{hetodpalib}(colour online) Variation in He appm production to dpa as a function of element and choice of nuclear data library. The data shown in the plot is also available in tabulated form in the supplementary data.
}
\end{figure}

Figures~\ref{heproductionlib} and~\ref{hetodpalib} show, respectively, the He production rate during 1~fpy and He-to-dpa ratios for the selected elements under DEMO conditions and as a function of the four nuclear libraries. Both figures show interesting variation for certain elements, most notably for C, where JEFF-3.2 predicts He production  that is more than an order of magnitude lower than the others, but then the He-to-dpa ratio from the same library is actually about twice as high as \(\sim\)30~He~appm/dpa obtained using the other libraries.
On the other hand, for Be, where JEFF-3.2 similarly underestimates He production compared to the other libraries, the underestimate remains in the He-to-dpa ratio. This illustrates the difficulty  in assessing the quality of one library against another, especially when looking for a general purpose data source that is reliable for most (all) materials.

For some of the more important nuclear materials the variation is, fortunately, less pronounced. For example, the range of He appm production values in Fe are within 20~appm of a 130~appm average (TENDL-2014 is the most significant outlier in this case, predicting only 110~appm He) -- a variation of only 16\% --, while for H production the range
of values are within about 30~appm of a 650~average -- a variation of less than 5\%.

\begin{figure}[h]
{\includegraphics[width=1.0\textwidth]
{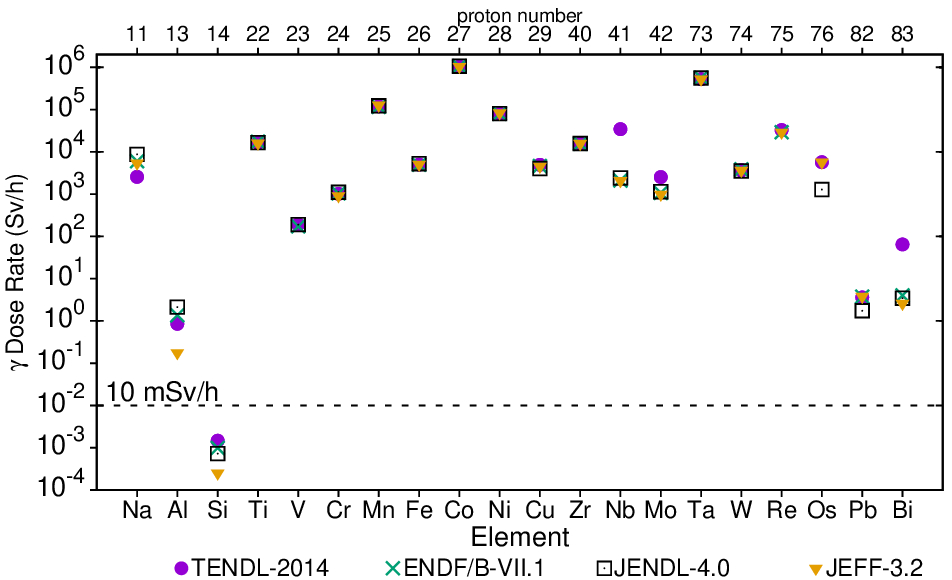}}
\caption{\label{doselib}(colour online) Variation in \(\gamma\)-dose rate from pure elements 12 days after shutdown following a 2~fpy DEMO-FW irradiation as a function of choice of nuclear data library. The data shown in the plot is also available in tabulated form in the supplementary data.}
\end{figure}

As mentioned in the introduction, activity predictions must be reliable because they will guide the limits for reactor operation, maintenance, and waste disposal. In particular, the time-decay in post-shutdown \(\gamma\)-dose rate from components determines when they can be handled -- either remotely, or by reactor operators. Figure~\ref{doselib} shows the \(\gamma\)-dose rate from the selected elements 12 days after shutdown following 2~fpy DEMO-FW irradiations for the different nuclear libraries. The three light elements Li, Be, and C do not appear in the plot because their irradiation does not produce any significant \(\gamma\)-emitting nuclides. At first glance, the variation doesn't appear too great, and certainly is not for important elements such as Fe and W. However, not all of the libraries have results for all of the elements. For Re, for example, which has potential applications in fusion as part of W-alloys, there is no value plotted for JENDL-4.0 because if falls below the \(1\times10^{-4}\)~Sv~hr\(^{-1}\) cutoff used in the figure (the supplementary data confirms that the value for JENDL-4.0 is \(6.0\times10^{-9}\)). Similarly, there is no result for V with JEFF-3.2 because the value is only \(2.0\times10^{-12}\)~Sv~hr\(^{-1}\). In both cases the missing values are many orders of magnitude different from values predicted with the remaining libraries.

Note that some of the differences shown in figure~\ref{doselib} may, in fact, be due to differences in the decay data used with each of the respective cross section libraries. ENDF/B-VII.1 has its own dedicated decay library, the JEFF-3.2 calculations were performed with a decay library released with JEFF-3.11, while JENDL-4.0, which is primarily aimed at fission fuel calculations, does not include a decay library covering the required radionuclides and so, as with TENDL-2014, UKDD-12 was used for those calculations.

As a final observation, note the 10~mSv~hr\(^{-1}\) level highlighted in figure~\ref{doselib}, which is the typical remote recycling limit expected for fusion reactor components (see, for example, \cite{ehrlich1999,elguelbaly2005}). From the plot we can see that only Si is predicted to be below the limit after 12 days of cooling.

\section{Summary}

A comprehensive scoping study has been performed to evaluate the activation, transmutation (depletion and build-up), and primary damage response under neutron irradiation for all naturally occurring elements from hydrogen to bismuth. The goal of this work is to provide a database of responses, covering the majority of nuclear technological space, that can be used to guide the selection, by engineers and materials specialists, of material compositions for nuclear applications. Special focus is given to the in-vessel conditions predicted for a future demonstration fusion power plant (DEMO), where many of the material choices have yet to be made. Results are presented from simulations with the inventory code FISPACT-II~\cite{subletetal2015} using the latest version of the modern TENDL nuclear data libraries --– TENDL-2014~\cite{tendl2014}.

The automated computational methodology developed to enable such a large project to be tackled on a reasonable timescale, which involves tens of thousands of inventory simulations, and a variety of post-processing techniques, is modern and efficient, and, in particular, can be easily applied to other irradiation scenarios and other nuclear data sets. The infrastructure has already been applied to typical fission environments and three other international nuclear data libraries.

Analysis of the differences in gas production, damage accumulation (measured via dpa), and transmutation burn-up between fusion and fission highlights that material predictions based on the fission experience might not be a reliable guide to performance and lifetime of similar materials in a fusion reactor. Furthermore, the results suggest that experimental campaigns in test fission reactors, such as HFR, will not be able to match the rates of material change expected in DEMO, confirming that a dedicated fusion materials test facility might be the only option.

Calculations with four of the most modern and complete nuclear data libraries, including, TENDL-2014, are in relatively good agreement for many important materials. However, there are still significant differences (especially for materials not selected for analysis in the present work), which demonstrates the need for careful selection of libraries for particular applications, and the need to perform detailed validation exercises for libraries. Similar automated calculations are now being routinely applied to the validation and verification of the latest TENDL libraries as they are released (see for example~\cite{fns2015,integral2015}).

\section{Acknowledgements}

 This work was partially carried out within the framework of the EUROfusion Consortium and has received funding from the Euratom research and training programme 2014-2018 under grant agreement number 633053 and from the RCUK Energy Programme [grant number EP/I501045]. To obtain further information on the data and models underlying this paper please contact PublicationsManager@ccfe.ac.uk. The views and opinions expressed herein do not necessarily reflect those of the European Commission.

\bibliographystyle{elsarticle-num}
\bibliography{IAEA_paper}

\begin{thebibliography}{10}
\expandafter\ifx\csname url\endcsname\relax
  \def\url#1{\texttt{#1}}\fi
\expandafter\ifx\csname urlprefix\endcsname\relax\def\urlprefix{URL }\fi
\expandafter\ifx\csname href\endcsname\relax
  \def\href#1#2{#2} \def\path#1{#1}\fi

\bibitem{mcnp1}
J.~T. Goorley, {MCNP6.1.1-Beta} release notes, {Los Alamos document number:
  LA-UR-14-24680}. Further details at \url{http://mcnp.lanl.gov} (2014).

\bibitem{subletetal2015}
J.~{\relax -Ch}. Sublet, J.~W. Eastwood, J.~G. Morgan, M.~Fleming, M.~R.
  Gilbert, \href{http://www.ccfe.ac.uk/fispact.aspx}{The {FISPACT-II} {U}ser
  {M}anual}, Tech. Rep. CCFE-R(11)11 Issue 7 (2015).
\newline\urlprefix\url{http://www.ccfe.ac.uk/fispact.aspx}

\bibitem{gilbertmariansublet2015}
M.~R. Gilbert, J.~Marian, J.~{\relax -Ch}. Sublet, Energy spectra of primary
  knock-on atoms under neutron irradiation, J. Nucl. Mater 467 (2015) 121--134.
\newblock \href {http://dx.doi.org/10.1016/j.jnucmat.2015.09.023}
  {\path{doi:10.1016/j.jnucmat.2015.09.023}}.

\bibitem{tendl2014}
A.~J. Koning, D.~Rochman, S.~C. van~der Marck, J.~Kopecky, J.-{\relax Ch}.
  Sublet, S.~Pomp, H.~Sjostrand, R.~A. Forrest, E.~Bauge, H.~Henriksson,
  O.~Cabellos, S.~Goriely, J.~Leppanen, H.~Leeb, A.~Plompen, R.~Mills,
  S.~Hilaire, {TENDL-2014}, available from
  \url{ftp://ftp.nrg.eu/pub/www/talys/tendl2014/tendl2014.html} (Release Date:
  December 11, 2014.).

\bibitem{forrest2007}
R.~A. Forrest, {The European Activation File: EAF-2007 decay data library},
  Tech. Rep. UKAEA FUS 537, EURATOM/CCFE, available from
  \url{http://www.ccfe.ac.uk/EASY2007.aspx} (2007).

\bibitem{fusionhandbook2015}
M.~R. Gilbert, J.-{\relax Ch}. Sublet, R.~A. Forrest, {Handbook of activation,
  transmutation, and radiation damage properties of the elements simulated
  using FISPACT-II \& TENDL-2014; Magnetic Fusion Plants}, Tech. Rep.
  CCFE-R(15)26, CCFE, available to download from
  {\url{http://www.ccfe.ac.uk/fispact{\textunderscore}handbooks.aspx}} (2015).

\bibitem{endfb72011}
\relax{M. B. Chadwick et al.},
  \href{http://www.nndc.bnl.gov/endf/b7.1/}{{ENDF/B-VII.1} nuclear data for
  science and technology: Cross sections, covariances, fission product yields
  and decay data}, Nuclear Data Sheets 112 (2011) 2887--2996.
\newblock \href {http://dx.doi.org/10.1016/j.nds.2011.11.002}
  {\path{doi:10.1016/j.nds.2011.11.002}}.
\newline\urlprefix\url{http://www.nndc.bnl.gov/endf/b7.1/}

\bibitem{JEFF32}
{T}he {JEFF}~team, \href{http://www.oecd-nea.org/dbdata/jeff}{{JEFF-3.2}:
  Evaluated nuclear data library} (2014).
\newline\urlprefix\url{http://www.oecd-nea.org/dbdata/jeff}

\bibitem{jendl4.0}
K.~Shibata, O.~Iwamoto, T.~Nakagawa, N.~Iwamoto, A.~Ichihara, S.~Kunieda,
  S.~Chiba, K.~Furutaka, N.~Otuka, T.~Ohsawa, T.~Murata, H.~Matsunobu,
  A.~Zukeran, S.~Kamada, J.~Katakura,
  \href{http://wwwndc.jaea.go.jp/jendl/j40/j40.html}{{JENDL}-4.0: A new library
  for nuclear science and engineering}, Nucl. Sci. Technol. 48~(1) (2011)
  1--30.
\newblock \href {http://dx.doi.org/10.1080/18811248.2011.9711675}
  {\path{doi:10.1080/18811248.2011.9711675}}.
\newline\urlprefix\url{http://wwwndc.jaea.go.jp/jendl/j40/j40.html}

\bibitem{pwrfissionhandbook}
M.~R. Gilbert, J.-{\relax Ch}. Sublet, {Handbook of activation, transmutation,
  and radiation damage properties of the elements simulated using FISPACT-II \&
  TENDL-2014; Nuclear Fission plants (PWR focus)}, Tech. Rep. UKAEA-R(15)31,
  UKAEA, available to download from
  \url{http://www.ccfe.ac.uk/fispact{\textunderscore}handbooks.aspx} (2015).

\bibitem{hfrfissionhandbook}
M.~R. Gilbert, J.-{\relax Ch}. Sublet, {Handbook of activation, transmutation,
  and radiation damage properties of the elements simulated using FISPACT-II \&
  TENDL-2014; Nuclear Fission plants (HFR focus)}, Tech. Rep. UKAEA-R(15)32,
  UKAEA, available to download from
  \url{http://www.ccfe.ac.uk/fispact{\textunderscore}handbooks.aspx} (2015).

\bibitem{fbrfissionhandbook}
M.~R. Gilbert, J.-{\relax Ch}. Sublet, {Handbook of activation, transmutation,
  and radiation damage properties of the elements simulated using FISPACT-II \&
  TENDL-2014; Nuclear Fission plants (FBR focus)}, Tech. Rep. UKAEA-R(15)33,
  UKAEA, available to download from
  \url{http://www.ccfe.ac.uk/fispact{\textunderscore}handbooks.aspx} (2015).

\bibitem{forrest1998}
R.~A. Forrest, Importance diagrams --- a novel presentation of the response of
  a material to neutron irradiation, Fus. Eng. Des. 43 (1998) 209--235,
  \url{http://dx.doi.org/10.1016/S0920-3796(98)00418-9}.

\bibitem{gilbertetalNSE2014}
M.~R. Gilbert, L.~W. Packer, J.~{\relax -Ch}. Sublet, R.~A. Forrest, Inventory
  simulations under neutron irradiation: Visualization techniques as an aid to
  materials design, Nucl. Sci. Eng. 177 (2014) 291--306.

\bibitem{norgettetal1975}
M.~J. Norgett, M.~T. Robinson, I.~M. Torrens, A proposed method of calculating
  displacement dose rates., Nucl. Eng. Des. 33 (1975) 50--54.
\newblock \href {http://dx.doi.org/10.1016/0029-5493(75)90035-7}
  {\path{doi:10.1016/0029-5493(75)90035-7}}.

\bibitem{greenwoodsmither1981}
L.~R. Greenwood, R.~K. Smither, {Displacement Damage Calculations with
  ENDF/B-V}, IAEA Vienna, Austria, 1981, pp. 185--192, in Proceedings of the
  Advisory Group Meeting on Nuclear Data for Radiation Damage Assessment and
  Reactor Safety Aspects, October 12-16.

\bibitem{macfarlanekahler2010}
R.~E. MacFarlane, A.~C. Kahler, Methods for processing {ENDF/B-VII} with
  {NJOY}, Nucl. Data Sheets 111 (2010) 2739--2890.
\newblock \href {http://dx.doi.org/10.1016/j.nds.2010.11.001}
  {\path{doi:10.1016/j.nds.2010.11.001}}.

\bibitem{gilbertsublet2011}
M.~R. Gilbert, J.-{\relax Ch}. Sublet, Neutron-induced transmutation effects in
  {W} and {W}-alloys in a fusion environment, Nucl. Fus. 51 (2011) 043005,
  \url{http://dx.doi.org/10.1088/0029-5515/51/4/043005}.

\bibitem{endf605}
M.~Herman, A.~Trkov (Eds.), {ENDF-6 Formats Manual, Data Formats and Procedures
  for the Evaluated Nuclear Data File ENDF/B-VI and ENDF/B-VII}, Vol.
  BNL-90365-2009 Rev.\ 2, Brookhaven National Laboratory, 2011,
  \url{https://www-nds.iaea.org/exfor/x4guide/manuals/endf-manual-v7.pdf}.

\bibitem{ehrlich1999}
K.~Ehrlich, The development of structural materials for fusion reactors, Phil.
  Trans. R. Soc. Lond. A 357 (1999) 595--623,
  \url{http://www.jstor.org/stable/55129}.

\bibitem{elguelbaly2005}
L.~El-Guebaly, P.~Wilson, M.~Sawan, D.~Henderson, A.~Varuttamaseni, Recycling
  issues facing target and {RTL} materials of inertial fusion designs, Nucl.
  Instr. Meth. Phys. Res. A 544 (2005) 104--110,
  \url{http://dx.doi.org/10.1016/j.nima.2005.01.199}.

\bibitem{fns2015}
J.-{\relax Ch}. Sublet, M.~R. Gilbert, {Decay heat validation, FISPACT-II \&
  TENDL-2014, JEFF-3.2, ENDF/B-VII.1 and JENDL-4.0 nuclear data libraries},
  Tech. Rep. {CCFE-R(15) 25}, CCFE, available from
  \url{http://www.ccfe.ac.uk/fispact{\textunderscore}validation.aspx} (2015).

\bibitem{integral2015}
M.~Fleming, J.-{\relax Ch}. Sublet, J.~Kopecky, {Integro-differential
  Verification and Validation, FISPACT-II \& TENDL-2014 nuclear data
  libraries}, Tech. Rep. {CCFE-R(15) 27}, CCFE, available from
  \url{http://www.ccfe.ac.uk/fispact{\textunderscore}validation.aspx} (2015).

\end{thebibliography}

\end{document}